\documentclass[conference]{IEEEtran}
\IEEEoverridecommandlockouts
\usepackage{cite}
\usepackage{amsmath,amssymb,amsfonts}
\usepackage{algorithmic}
\usepackage{graphicx}
\usepackage{textcomp}
\usepackage{xcolor}
\def\BibTeX{{\rm B\kern-.05em{\sc i\kern-.025em b}\kern-.08em
    T\kern-.1667em\lower.7ex\hbox{E}\kern-.125emX}}
\begin{document}


\title{A real-time rendering method for high albedo anisotropic materials with multiple scattering
\thanks{This work was supported by the Beijing Municipal Science and Technology Commission and the Zhongguancun Science and Technology Park Business Management Committee, as well as Department of Housing and Urban-Rural Development of Guangdong Province, with project numbers Z221100007722005, 20230468276, and 2023k38232558.}
}

\author{\IEEEauthorblockN{1\textsuperscript{st} Shun Fang}
\IEEEauthorblockA{\textit{Lumverse Reserch Institute} \\
\textit{Beijing Lumverse Technology Co., Ltd}\\
Beijing, China \\
fangshun@ieee.org}
\and
\IEEEauthorblockN{2\textsuperscript{nd} Xing Feng}
\IEEEauthorblockA{\textit{Lumverse Reserch Institute} \\
\textit{Beijing Lumverse Technology Co., Ltd}\\
Beijing, China \\
fengxing@lumverse.com}
\and
\IEEEauthorblockN{3\textsuperscript{rd} Ming Cui}
\IEEEauthorblockA{\textit{Lumverse Reserch Institute} \\
\textit{Beijing Lumverse Technology Co., Ltd}\\
Beijing, China \\
cuiming@lumverse.com}
}

\maketitle

\begin{abstract}
We propose a neural network-based real-time volume rendering method for realistic and efficient rendering of volumetric media. The traditional volume rendering method uses path tracing to solve the radiation transfer equation, which requires a huge amount of calculation and cannot achieve real-time rendering. Therefore, this paper uses neural networks to simulate the iterative integration process of solving the radiative transfer equation to speed up the volume rendering of volume media. Specifically, the paper first performs data processing on the volume medium to generate a variety of sampling features, including density features, transmittance features and phase features. The hierarchical transmittance fields are fed into a 3D-CNN network to compute more important transmittance features. Secondly, the diffuse reflection sampling template and the highlight sampling template are used to layer the three types of sampling features into the network. This method can pay more attention to light scattering, highlights and shadows, and then select important channel features through the attention module. Finally, the scattering distribution of the center points of all sampling templates is predicted through the backbone neural network. This method can achieve realistic volumetric media rendering effects and greatly increase the rendering speed while maintaining rendering quality, which is of great significance for real-time rendering applications. Experimental results indicate that our method outperforms previous methods.
\end{abstract}


\section{Introduction}
Volume media are also called participating media. When light passes through volume media such as clouds, smoke, fog, dust, jade, milk, skin, wax, fruit pulp, etc., it will produce tens of thousands of refraction, scattering, absorption and other phenomena. How to render volumetric media realistically and efficiently has always puzzled academia and industry. Such questions have been the focus of numerous investigations, involving light transport in general, or specifically for rendering-participating media (such as photon beams19\cite{b1}, photon surfaces\cite{b2}, and volumetric path guidance\cite{b3}). However, it takes a long time for light to converge to noise-free conditions in a volume medium. In recent years, researchers have been making breakthroughs in the hope of finding better methods, including diffusion theory and Monte Carlo integration. However, significant performance improvements were achieved from Kallweit et al\cite{b4} using radiation prediction neural networks (RPNN) and showed how to use neural networks(NN)to estimate scattered irradiance.

Although it has achieved good results, the RPNN architecture hinders its own performance. It only accepts density features as input, requires sampling a large number of features to obtain good results, and cannot handle the shadow part of the medium well, so it is difficult for the network to infer the basic principles of optical transmission. Principles of physics.

In order to obtain the same rendering quality with less calculation, this article creates a network architecture that can better approximate the solution of RTE, and uses a neural network to simulate the iterative integration process of solving the radiation transfer equation, thereby accelerating volume rendering of volume media. speed. We change the phenomenon of only inputting density features and separately sample the density features, transmission features and phase features of the volume medium. This sampling method can simplify the network structure and input parameters and achieve faster reasoning. In addition, instead of sending all three types of decomposed features into the network, we use diffuse reflection templates and highlight templates to divide 12 layers, and send the sampled features of each layer into the network for internal scattering estimation, which can generate real volume rendering in real time.

\section{Related Work}
The Monte Carlo method is a general term for a class of algorithms that solve problems through random sampling. The problem to be solved is the probability of a random event or the expectation of a random variable. Through the random sampling method, the probability of a random event is estimated based on its frequency and used as the solution to the problem. Solving RTE, i.e. finding all potential light paths connecting light sources, cameras and medium vertices, is at the core of these methods. Kajiya and Von Herzen\cite{b5} were the first to use path tracing to numerically estimate radiative transfer in volumes. Their technique was later extended bidirectionally, by constructing bidirectional path tracing\cite{b6}, mutating paths using the Metropolitan-Hastings algorithm\cite{b7}, and the importance of low-order scattering Sexual sampling\cite{b8,b9}. The efficiency of these algorithms can be improved by computational and related estimates such as radiosity caching[1], multi-light rendering\cite{b10,b11} and density estimation based methods\cite{b12,b13}. Krivanek et al\cite{b14}. formulated a unified theory for these seemingly incompatible methods, allowing their advantages to be combined into a single robust estimator. Bitterli and Jarosz\cite{b15} Generalized volumetric density estimation for arbitrary dimensional samples. However, even with advanced methods for constructing free paths and transmittance estimates\cite{b16} operating on special data structures\cite{b17,b18}, when using our target These methods fall far short of interactive frame rates for highly scattering materials. Although these methods are unbiased and they all speed up the rendering process significantly, they are still time-consuming and can only be used for offline rendering\cite{b4}.

In addition to MC integrators, there is another type of RTE solver used to efficiently capture multi-scattered volume illumination\cite{b19}. That is, diffusion theory is used to approximate multiple scattering, with the purpose of solving the polydispersion problem of MC integrators. In Stam, he represented inhomogeneous media by solving the discrete form of the diffusion equation on a grid. Koerner et al.\cite{b20} adjust the diffusion coefficient to improve accuracy in areas of low density and/or high albedo. Several semi-analytical solutions based on combining the transport contributions of two or more monopoles have been developed to rapidly simulate subsurface scattering\cite{b21,b22}. Diffusion-based methods\cite{b23} solve higher-order problems, but they are hampered by other problems that prevent them from being used in real time. For example, their computational cost depends polynomially on the grid resolution, which is also constant. However, in cases of insufficient resolution, deviations can be distinguished. Since they do not support progressive rendering, it is difficult to distribute the computational load between frames. Furthermore, none of the diffusion methods introduced above handles highly anisotropic scattering in clouds well, and appearances from low- and medium-order scattering, such as silver lines, are not well reproduced.

Neural Networks. Deep neural networks (see Bengio et al.\cite{b24}, LeCun et al.\cite{b25} for comprehensive reviews) can effectively model complex relationships between input and output variables in a highly nonlinear manner. This data-driven approach has become state-of-the-art for a variety of challenging problems, such as image recognition\cite{b26,b27}, machine translation\cite{b28}, or raw audio and Generative modeling of natural images\cite{b29,b30}. Deep learning has also been successfully applied to computer graphics problems. Nalbach et al.\cite{b31} use convolutional neural networks (CNNs) to synthesize ambient occlusion, lighting, and other effects in screen space. Bako et al.\cite{b32} and Chaitanya et al.\cite{b33} use neural networks to denoise rendered images. These methods operate in a two-dimensional environment, using color images and feature images as input to the network. Similar to Chu and Thuerey\cite{b34}, who applied CNNs to fluid simulations, our network operates in 3D. Neural networks have also been used to predict the exit point position of objects from internal scattering\cite{b35}, or directly estimate the contribution of all possible paths between two specific points in homogeneous media\cite{b36}. Mildenhall et al.\cite{b37} utilize neural radiation fields to represent scenes and render them via ray travel. The neural radiation field is further used to accelerate the convergence speed of path tracking\cite{b38}. These methods provide valuable insights into the application of learning techniques for rendering purposes.

Radiation prediction neural network. Researchers have been looking for approximate but more effective MLP-based methods to achieve faster performance. Research starting from Kallweit et al.[4] has attracted our attention. They were the first to introduce neural networks to estimate the challenging inscatter term and were able to synthesize images of clouds in minutes by successfully avoiding the time-consuming process of tracing a large number of light paths. The performance of this method is further improved by caching latent vectors\cite{b39}. Essentially, radiation prediction neural networks consider light transmission as a mapping from density fields to radiation fields. Since the mapping is very complex, the neural network must be large enough to approximate it, which makes the evaluation very time-consuming. However, traveling with light is easy. The task of the network can be simplified by using simply computed features as additional inputs. The third section of the article will describe the proposed network framework and specific real-time details in detail. With appropriate decomposition of input features and customized neural network structure, convergence can be achieved with very limited training data.


\section{Research contents}

\subsection{Overall processs}\label{AA}
The 3D volume medium first outputs three kinds of sampling features through the data processing module, and then input them to the feature extraction module (or the whole input) according to the sampling template, and finally predicts the scattering distribution of the central points of all the sampling templates.
\begin{figure}[htbp]
\centerline{\includegraphics[width=9cm,height=2cm]{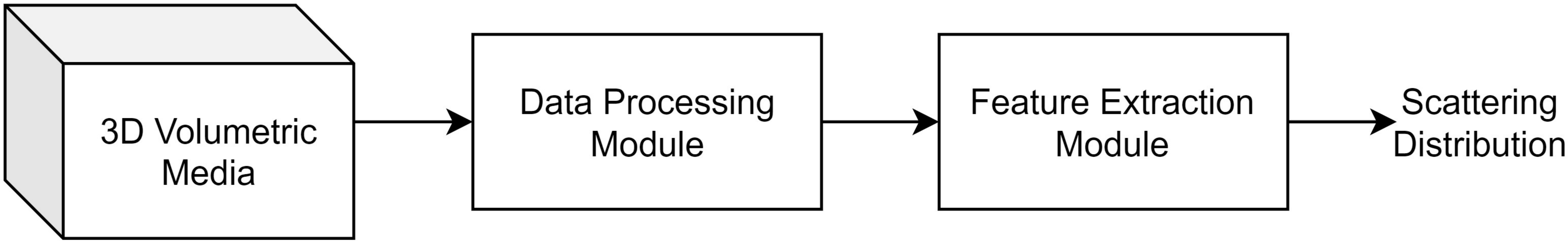}}
\caption{Overall flow chart.}
\label{fig1}
\end{figure}

\subsection{Data processing module}
First, the density field of 3D bulk medium is obtained, and the original density field is downsampled to obtain the density fields of different resolution levels, and then input into the transmission field formula to obtain the graded transmission field, and all the graded transmission fields get the transmission features through 3D-CNN.3D volumetric media, density field, and transmission features were sampled using sampling templates to obtain sampled phase features, sampling density features, and sampled transmission features.
\begin{figure}[htbp]
\centerline{\includegraphics[width=9cm,height=3cm]{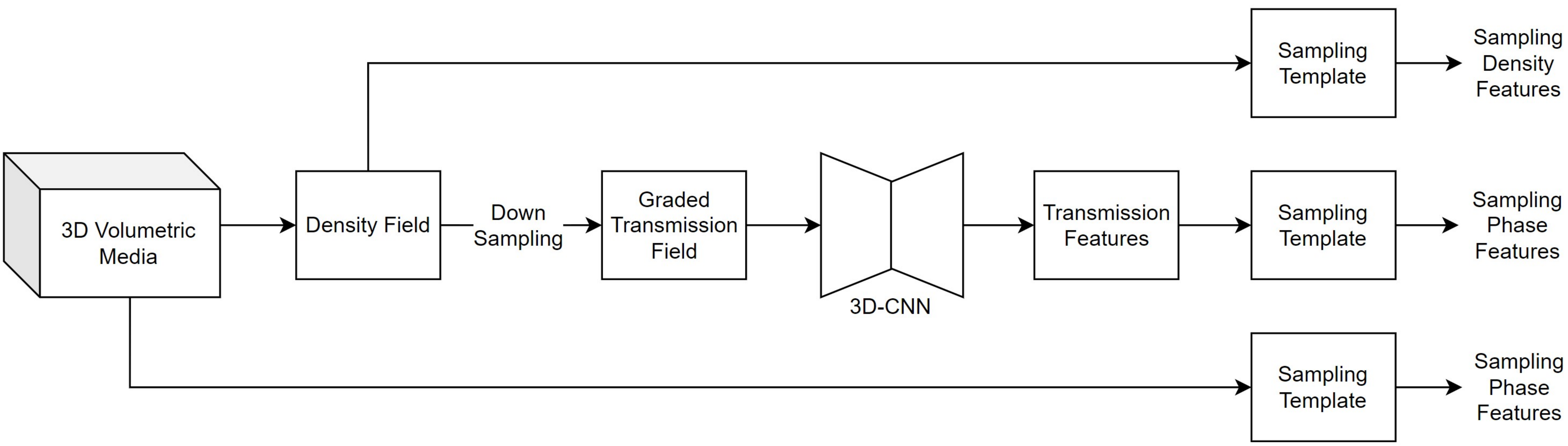}}
\caption{Data processing module flow chart.}
\label{fig2}
\end{figure}

\subsubsection{Transmission features}
The bulk media is represented by point cloud or voxel with the original resolution of 1024 * 1024 * 1024. We downsampled the density field to produce 11 density fields of 1024 * 1024 * 1024 * 1024, 512 * 512 * 512, 256 * 256 * 256, 128 * 128 * 128, 624 * 64 * 64, 32 * 32 * 32, 16 * 16 * 16, 8 * 8 * 8, 4 * 4 * 4 * 4, 2 * 2 * 2, 1 * 1 * 1.
The density fields of 11 resolution levels are respectively input into the following formula, yielding 11 levels of transmission fields, called graded transmission fields. The formula is as follows. ${T}_i$ is the transmission field represented by the density field,  $\rho_i(x)$ represents the i-th density level,  $\lambda^{i+1}$ is the coefficient in the range of 0-1, actual = 0.6.
\begin{equation}
\overline{T}_i(p,y)=e^{-\int_p^y\lambda^{i+1}\rho_i(x)dx}
\end{equation}

\begin{figure}[htbp]
\centerline{\includegraphics[width=8.5cm,height=3cm]{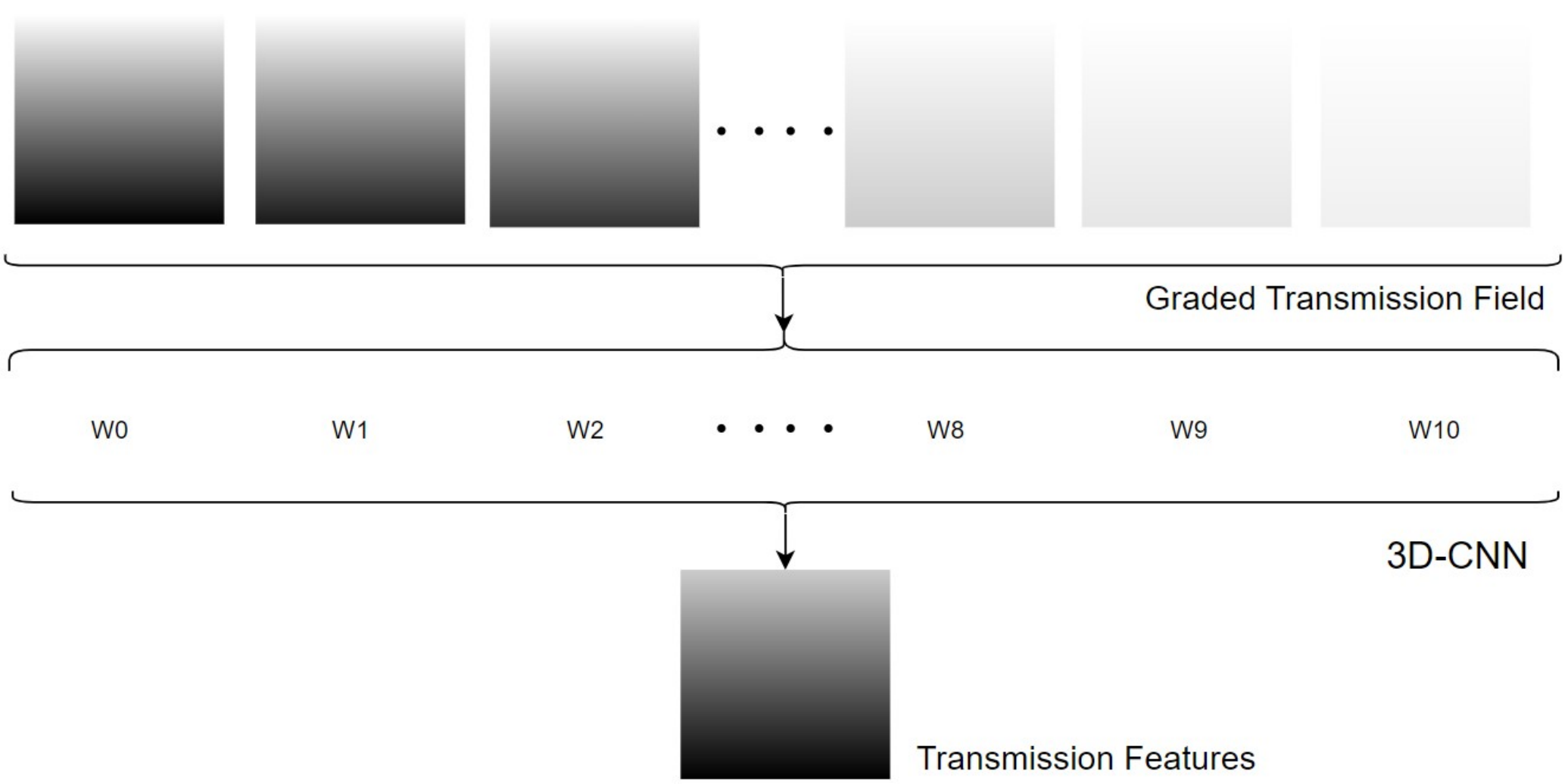}}
\caption{3D-CNN generates transmission fields.}
\label{fig3}
\end{figure}

Inputting 11 graded transmission fields into a 3D convolutional neural network (3D-CNN) is equivalent to adding a weight to each level of transmission field, and finally superimposes it into a transmission feature. As shown below Fig~\ref{fig3}.

\subsubsection{Sampling template}
Diffuse reflection sampling template. The first sampling template is the diffuse reflection sampling template, which uses a sphere as the sampling template. The advantage is that it can sample the scattering area and pay more attention to the scattering of light. The distance from the sampling center point will affect the scattering contribution. The farther away, the lower the contribution. Therefore, the diffuse reflection sampling template is divided into 8 layers according to distance:
\begin{equation}
S=\{s_1,s_2,...,s_n\}
\end{equation}

$S$ represents the template, which represents the $i-th$ template point, $n=8$, that is, a total of 8 layers of templates. The diffuse reflection template has a total of 8 layers and 182 template points. Each template point is evenly distributed. The farther the distance, the larger the volume of the ball, so more template points are needed. There are 6 sampling points in layer 0, of which 1 is at the center of the sphere and 5 are evenly distributed inside the sphere; layers 1 to 7 are 8, 12, 16, 24, 32, 48, and 48 respectively. sampling points, evenly distributed in the area between the surface of the current layer's sphere and the surface of the previous layer's sphere.

Highlight sampling template. The second sampling template is the highlight sampling template, which samples along the direction of the incident light. Its advantage is that it can play an important role in the formation of highlights and shadows. From the time when the incident light penetrates the volume medium to the volume medium particles, it is divided into 4 layers evenly according to the distance. The number of template points is 32, 16, 16, and 8 respectively, for a total of 72 template points.

Diffuse reflection sampling template point uniform distribution algorithm. In order to distribute the template points evenly, an algorithm is used:
\begin{equation}
\mathcal{D}(S_i)=\frac1{N_i}\sum_{s\in S_i\{\varepsilon\}}\varepsilon\in S_imin\{\parallel\varepsilon-s\parallel\}
\end{equation}

$S_i$ is the i-th layer template, and  ${D}(S_i)$ is the average of the minimum values between all two template points, ${s\in S_i\{\varepsilon\}}$ means that  is the template point except $\varepsilon$ in the i-th layer template. It is assumed here that the radius of the sphere and the cone is unit length.
\begin{equation}
r_i=\frac{2^{i-1}}{2^8\times\mathcal{D}(S_i)}\xi_i
\end{equation}

$r_i$ represents the radius of the i-th layer template, and ${D}(S_i)$ is the shortest distance under unit radius, so it must be much less than $l$. $2^8$ is the resolution of the first-level mipmap, and $\xi_i=1-0.08min(i,5)$ is the empirical parameter that controls the template overlap rate. The smaller  is, the smaller the radius is.
 
\subsubsection{Phase function}
Phase function of a homogeneous medium. For a uniform volume medium, its phase function has nothing to do with position $p$. At this time $\mu_{a}$, $\mu_{s}$ and $\mu_{t}$ are constants, where $\mu_{a}$ is the absorption coefficient, $\mu_{s}$ is the scattering coefficient, $\mu_{t}=\mu_{a}+\mu_{s}$ is the decay coefficient, which describes the features of radiation attenuation in the medium. At this time, the phase function is a one-dimensional function: $f_p(\theta)$, the $\theta$ are called scattering angles, which are the angles between the incident direction of the light and the emergent direction after scattering. This article uses the Henyey-Greenstein Phase Function (HG phase function for short), which is widely used to describe uniform volume media such as smoke, clouds, and fog. Its formula is as follows:
\begin{equation}
f_p(\theta)=\frac1{4\pi}\frac{1-g^2}{(1+g^2.2g\cos\theta)^{\frac32}}
\end{equation}

$g\in(-1,1)$ is called the asymmetry parameter, and its value is close to the average cosine of the light scattering angle in all directions. If $g>0$, then mainly forward scattering occurs, scattering more energy in the direction opposite to the incident direction. If $g<0$, then backscattering mainly occurs, with more energy scattered in the incident direction.
Multimodal phase function. Multi-Henyey-Greenstein Phase Function (MHG phase function for short) is a linear combination of two or more HG phase functions. Its formula is as follows:
\begin{equation}\int_{x}^{MHG}(\theta)=\lambda_1f_p(\theta,g_1)+\lambda_2f_p(\theta,g_2)+...+\lambda_Nf_p(\theta,g_N)\end{equation}

$f_p(\theta,g_i)$ is formula 3, $\lambda_1+\lambda_2+...+\lambda_N=1$, where $\lambda$ is the weight coefficient of each  is the number of , and $g_i$ indicates that each phase function has its own asymmetry parameter.
Isotropic phase function. The isotropic phase function means that after light is scattered in a volume medium, the probability of scattering in all directions is the same, so the phase function formula is:
\begin{equation}f_p(\theta)=\frac1{4\pi}\end{equation}
 
Phase functions of different materials. Divide 3D volume media into three major categories: gas, solid, and skin. Gaseous types include clouds, smoke, and fog; solid and liquid types include jade, milk, wax, and pulp; and skin types include human skin. Different material types have different forms and parameters of their phase functions. In this paper, the phase function of air adopts an isotropic phase function. The gas state adopts the  phase function; the solid and liquid adopts the bimodal  phase function, that is, two  phase functions; human skin adopts the multimodal  phase function, and this article uses three  phase functions.

\subsubsection{Phase features}
The phase feature uses sampling template points to calculate the phase distribution of medium particles and serves as an input parameter of the neural network to predict the real light scattering distribution. The formula for the phase features of any template point $s_i$ relative to point $p$ is as follows:
\begin{equation}f_i=f^*(\omega,s_i-p)\cdot f^*(s_i-p,l)\end{equation}

\begin{figure}[htbp]
\centerline{\includegraphics[width=8.5cm,height=2.8cm]{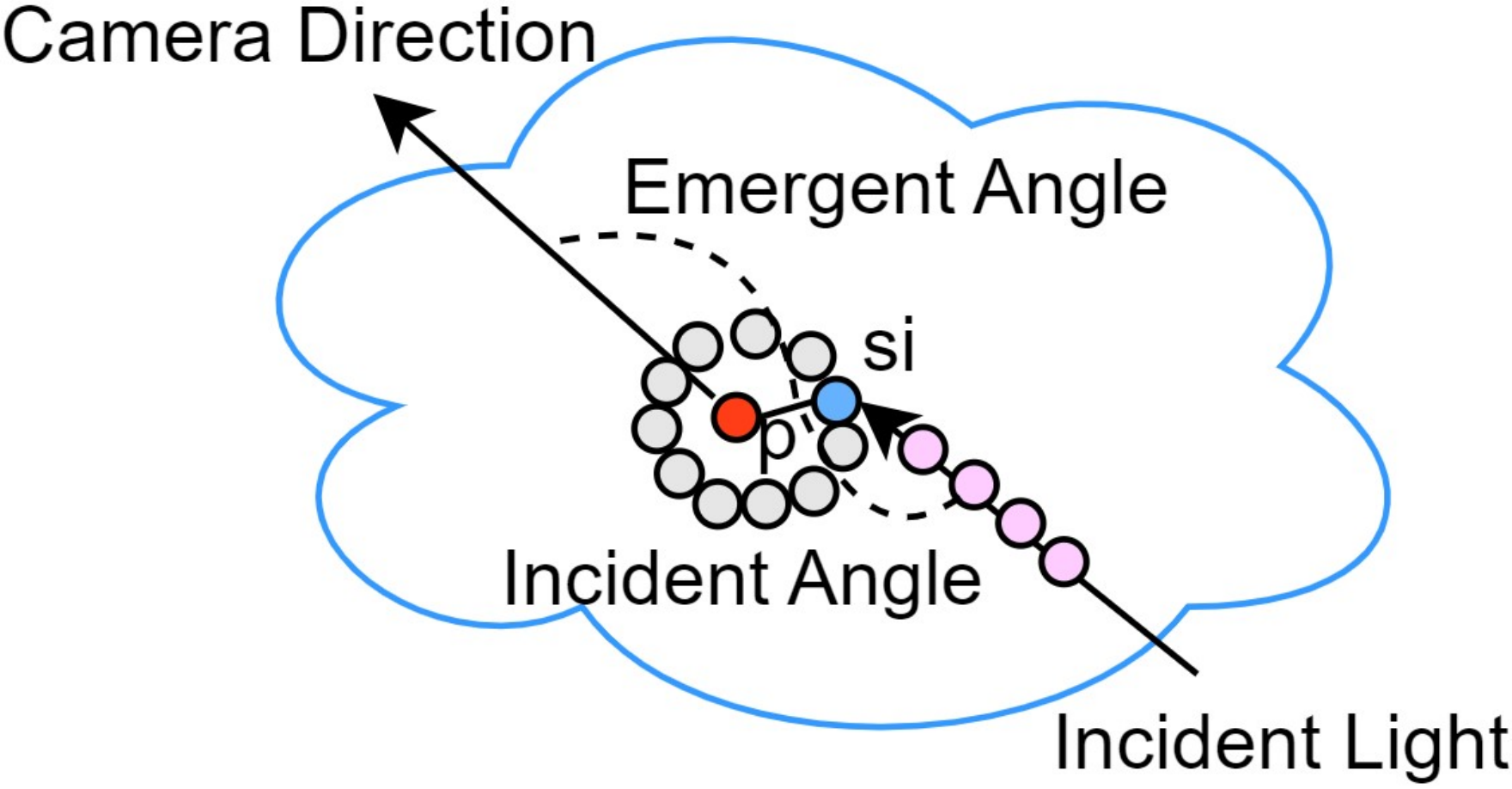}}
\caption{Light relationship of template point at position p.}
\label{fig4}
\end{figure}

$s_i$ is the i-th template point, $p$ is the volume medium particle position, $f_i$ is the phase value of the i-th template point, and $ f^*$ is the volume phase function. The incident light $l$ is the direction of light incidence, and the camera direction $\omega$ is also the direction of sight. $\omega,s_i-p$ is the emergent angle, $s_i-p,l$ is the incident angle, as shown in fig~\ref{fig4}.
The gray point is the diffuse reflection template point, and $si$ is one of the diffuse reflection template points. The red point $p$ is the medium particle position, and the pink point is the highlight template point.
Volume medium particles are infinitely small points, but template points are points of volume. Different positions of the incident light at the template point will affect the incident angle and emergent angle, as shown in the following fig~\ref{fig5}
\begin{figure}[htbp]
\centerline{\includegraphics[width=8.2cm,height=2.5cm]{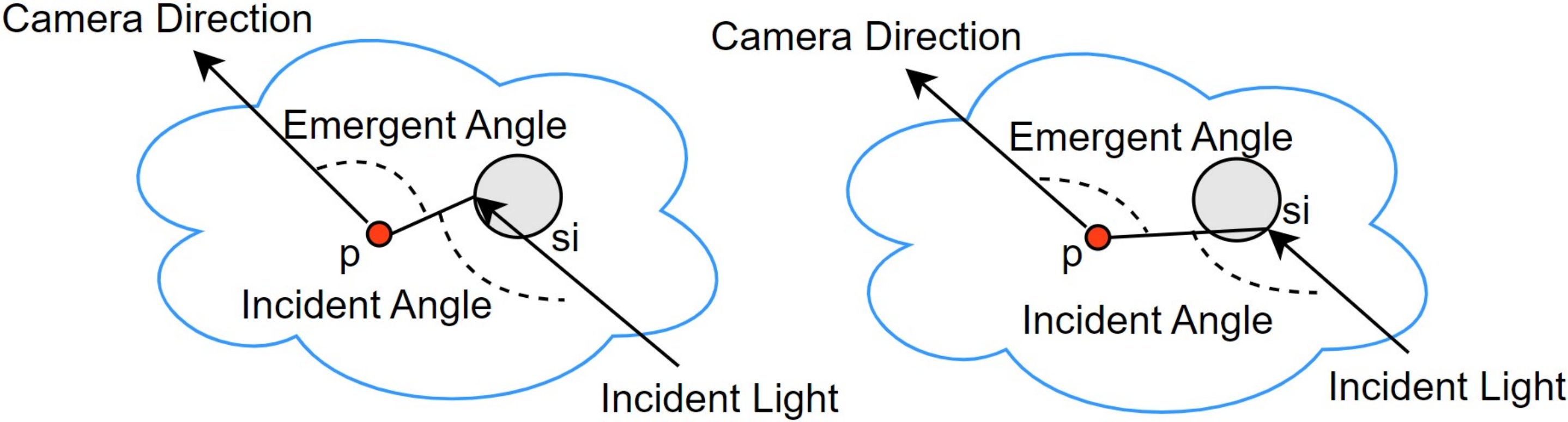}}
\caption{Template point volume influence.}
\label{fig5}
\end{figure}

In order to solve this problem, it is necessary to integrate along the volume of the template point, the formula is as follows:
\begin{equation}f_{\omega}^{*}=\int_{\Omega}f(\omega,\varphi)d\varphi,f_{l}^{*}=\int_{\Omega}f(\varphi,l)d\varphi \end{equation}

$\Omega=\{\varphi|(\theta\cdot\varphi)>\cos(^\varrho/_2)\}$, $f$ is the HG phase function, $\theta$ is the incident angle, $\varphi$ is the emergent angle, and  represents the solid angle projected from the current template point to point $p$. $f_{\omega}^{*}$ is the integral form of the phase function, which is called the volume phase in the camera direction. $f_{l}^{*}$ is also the integral form of the phase function, which is called the volume phase in the direction of the incident light. $\varphi$ is multiplied by $\theta$, including all angles greater than $\cos(^\varrho/_2)$, and $\Omega$ is the angle space composed of all $\varphi$.

\subsubsection{Precomputed sampling features}
Sampling features include sampling transmission features and sampling phase features. 4129 template samplings were performed for each volume of medium, using a uniform sampling strategy. Each time the template is sampled, 254 sampled transmission features and 254 phase transmission features will be generated.

Precompute sampling phase features. First find the volume phase of all template points through formula 6, and set all possible angles between the incident light and the camera direction, then find the phase features through formula 2, formula 3, formula 4 and formula 5  , and pre-calculate all sampling points Then, save it to the precalculated table.

Precomputed sampled transmission features. The graded density field is obtained by downsampling the density field to obtain the graded transmission field, and transmission features are generated through 3D-CNN. The sampling transmission features are obtained through the sampling template. All sampling points are pre-calculated and stored in the pre-calculation table.

\subsection{Feature extraction module}

\subsubsection{Radiation transfer equation}
The classical transmission equation is as follows.
\begin{equation}
\begin{split}
\frac\partial{\partial\omega}L_o(p,\omega)=L_e(p,\omega)-\mu_a(p)L_i(p,\omega)-\mu_s(p)L_i(p,\omega)+ \\
\mu_s(p)\int_{\Omega}f_p(\omega_i\to\omega)L_i(p,\omega_i)\mathrm{d}\omega_i
\end{split}
\end{equation}

\begin{equation}
\begin{split}
\frac\partial{\partial\omega}L_o(p,\omega)=L_e(p,\omega)-\mu_a(p)L_i(p,\omega)+ \\
\mu_s(p)\int_\Omega f_p(\omega_i\rightarrow\omega)L_i(p,\omega_i)\mathrm{d}\omega_i
\end{split}
\end{equation}

When light $L$ passes through the volume medium along the direction $\omega$ at position $p$, the subscript  of the change amount of radiation loudness $L_o$ is the initial light, and the subscript i of $L_i$ is the incident light. Point $p$ is a particle of a volume medium that can absorb and scatter incident light. $L_e(p,\omega)$ is the self-illumination term of medium particles, that is, the increment of medium self-luminescence to radiant brightness. $\mu_a(p)L_i(p,\omega)$ is the light absorbed by the particles (referred to as the absorption term), and $\mu_s(p)L_i(p,\omega)$ is the light scattered by the particles to other directions. $\mu_t(p)L_i(p,\omega){=\mu_a(p)L_i(p,\omega)+\mu_s(p)L_i(p,\omega)}$. Albedo $\eta=frac\mu_s\mu_t$, mean free path $l=frac\mu_t$ is the average distance traveled by light during two consecutive interactions with particles in a medium. $\mu_s(p)\int_0f_p(\omega_i\to\omega)L_i(p,\omega_i)\mathrm{d}\omega_i$ is that the direction in which light is scattered by particles is the same as the current direction (referred to as the co-scattering term). $f_p(\omega_i\to\omega)$ is the scattering phase function from direction $\omega_i$ to $\omega$ at position $p$, that is, the ratio of the energy of light scattered from direction $\omega_i$ to $\omega$ to the energy scattered in all directions of $\Omega$, which describes the spatial distribution of scattered light at point $p$, $\Omega$ is a unit ball. Formula 10 represents the other direction and co-directional scattering terms separately. Formula 11 combines the two items together and is called the scattering term. All are calculated according to Formula 10.
\begin{equation}\begin{aligned}F(p,\omega)=\int_{\Omega}f_p(\omega_i\rightarrow\omega)L_i(p,\omega_i)\mathrm{d}\omega_i\end{aligned}\end{equation}

$F(p,\omega)$ is the co-scattering term (no scattering coefficient $\mu_s(p)$). Simplify the transmission equation, assuming that the volume medium does not emit light, the $L_e(p,\omega)$ terms can be ignored, and the medium albedo is assumed to be constant, then Equation 11 is simplified.
\begin{equation}\frac\partial{\partial\omega}L_o(p,\omega)=-\mu_t(p)L_i(p,\omega)+\mu_s(p)F(p,\omega)\end{equation}

The integral of both sides is:
\begin{equation}L_o(p,\omega)=T(p,z)L_i(z,\omega)+\int_p^zT(p,x)\mu_s(x)F(x,\omega)\mathrm{d}x\end{equation}

\begin{equation}T(p,y)=e^{-\int_p^y\mu_t(x)dx}\end{equation}

$\mathrm{z}=\lim_{x\to\infty}(p-x\cdot\omega)$ represents a reverse extreme point of the direction (that is, the direction $\omega$ is from $z$ to $p$). Assuming that the light source is far away (such as the sun) and ignoring other effective light sources, then:
\begin{equation}L_i(p,\omega)=\delta(<l,\omega>+1)\end{equation}

$\delta$ is the Dirac function, $I$ represents the intensity of light. Light only comes from direction $l$. Substituting Formula 14 into Formula 12, we get the follows.
\begin{equation}F(p,\omega)=\int_\Omega\left[\left(f_p\int_p^zT\mu_t\eta F\mathrm{d}x\right)+f_pTL_i\right]\mathrm{d}\omega_i\end{equation}

The input light of the next particle is the output light of the current particle, so $L_i(p,\omega)=L_o(p,\omega)$, so Formula 14 can be brought into Formula 12. Substituting Formula 16 into Formula 17, we get:
\begin{equation}F(p,\omega)=\eta\int_\Omega f_p\int_p^zT\mu_tF\mathrm{d}x\mathrm{d}\omega_i+f_pT\cdot I\end{equation}

$\int_\Omega f_pT\delta(\langle\iota,\omega\rangle+1)\cdot Id\omega_i=f_pT\cdot I\int_\Omega\delta(\langle\iota,\omega\rangle+1)d\omega_i,$, and the Dirac function $\int_0\delta(\langle\iota,\omega\rangle+1)d\omega_i=1$, so the above formula = $f_pT\cdot I$. Formula 18 there is $F$ on both sides of the equation, but the meaning is different. The $F$ on the left represents the co-scattering term of the next particle, while the  on the right represents the co-scattering term of the current particle. Let $\Pi F(p,\omega)=\int_\Omega f_p\int_p^zT\mu_tF(p,\omega)dxd\omega_i$, then $F_{i+1}(p,\omega)=\Pi F_i(p,\omega),F_0(p,\omega)=f_pT\cdot I.$
\begin{equation}F(p,\omega)=\sum_{i=0}^{\infty}\eta^{i}F_{i}(p,\omega)=F_{0}+\eta F_{1}+\eta^{2}F_{2}+\eta^{3}F_{3}+...\end{equation}

$F(p,\omega)$ is the sum of all scattering terms in the direction, given any approximate $F_{i}$, $F_{all}$ can be inferred through the neural network without complicated calculations. For co-directional scattering, its phase function $f_p(\omega_i\to\omega)$ has nothing to do with direction $\omega_i$ and is a constant. Let $\overline{\Pi}F(p,\omega)=\int_{\Omega}\int_p^zT\mu_tF(p,\omega)dx\mathrm{~}d\omega_i,\mathrm{~then~}F_{i+1}(p,\omega)=\Pi^*F_i(p,\omega),F_0(p,\omega)=T\cdot I$.

\begin{equation}F(p,\omega)=\sum_{i=0}^\infty\eta^if_iF_i(p,\omega)\end{equation}

\subsubsection{The backbone neural network}
The backbone neural network is shown in Figure 6. It should be noted that according to density, transmittance and phase, the backbone neural network in Figure 6 has three identical networks, but the input parameters and output parameters are different.

Input layer. The sampling features including sampling density features, sampling transmission features and sampling phase features, are pre-calculated and can be obtained directly from the pre-calculation table based on the position coordinates of particles and template points. The sampling features need to be input to the input layer of the network in batches, with a total of 4129 sampling templates, so they need to be input into the neural network 4129 times. In order to improve real-time rendering performance, multiple neural network models can be generated and calculated in parallel, thereby increasing the rendering frame rate.

"Configuration parameters" include the asymmetry parameters of the HG phase function, the albedo, and the angle between the incident light and the camera direction. Among them, the asymmetric parameters of the HG phase function are divided into one parameter, two parameters and three parameters according to the type of material. The materials of different volume media may have different reflection capabilities for light of different wavelengths, so we use a three-dimensional vector to represent the albedo, which respectively represents the albedo of a volume medium for red light, green light and blue light. Each The value range of the component is between 0 and 1. The value range of albedo varies according to the type of material. The value range of albedo of gaseous materials is 0 to 0.5, and the value range of albedo of other solid, liquid and skin materials is 0.5 to 1. In addition, we assume that the 3D volume medium is homogeneous, so the albedo within the same material is equal everywhere.

\begin{figure}[htbp]
\centerline{\includegraphics[width=8.5cm,height=2.5cm]{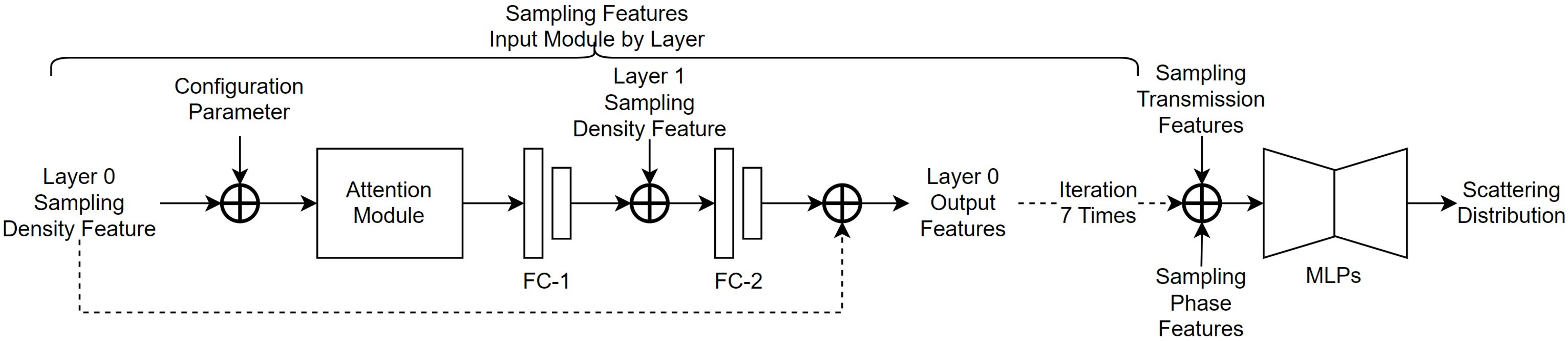}}
\caption{The structure of the backbone neural network.}
\label{fig6}
\end{figure}

Hidden layer. The sampling feature layer-by-layer input modules in Fig~\ref{fig6} have three identical network structures. Only the layer-by-layer input process for sampling density features is shown here. Other sampling transmission features and sampling phase features are all of the same structure.

The sampling density features of the current layer are fused with the "configuration parameters" and then input into the attention module. Its main purpose is to determine which channel features are more important through the attention mechanism. The feature vector output by the attention module is fused with the next layer of sampling density features through fully connected FC-1. The main purpose of FC-1 is to make the two feature vectors have the same dimension for subsequent fusion operations. The fused feature vector passes through the fully connected FC-2, and then is fused with the sampling density feature of the current layer and then output. Because the diffuse reflection sampling template has 8 layers and the highlight sampling template has 4 layers, this process needs to be carried out 12 times in total. It should be noted that the 8 layers of diffuse reflection and the 4 layers of specular reflection are not the same network. That is, the neural network is also divided into two neural networks with the same structure according to the sampling template. If density, transmittance and phase are also included, then the total There are 6 identical network modules. The diffuse reflection sampling density features after 8 iterations and the specular reflection sampling density features after 4 iterations are merged together through a full connection.

The sampling transmission features and sampling phase features are also input by layer 12 times according to the above process, and then fused together, and finally input into a multi-layer perceptron, and finally output a scattering distribution of the center point of the sampling template.

Output layer. All 4129 volume medium particle points are sampled with the template, and the above process is used to predict the output through the network in parallel, and finally the scattering distribution of the 4129 sampling template center points is obtained. These values are brought into the radiation transfer equation respectively, and the volume medium can be rendered. The scattering effect.
Loss function. Use the mean square error to calculate the loss function.
\begin{equation}
\mathcal{L}_S=\frac1{\mathcal{B}}\sum_{i\in\mathcal{B}}\left(log\left(\frac{\overline{T}_c^*(\mathcal{F})}{\eta_c^\gamma}+1\right)-log\left(\frac{F_i(p,l)}{\eta_c^\gamma}+1\right)\right)^2
\end{equation}

$\mathcal{B}$ is the Batch size, which is the number of samples used in one iteration of learning, $\overline{T}_c^*(\mathcal{F})$ is the estimated value of the neural network, ${F_i(p,l)}$ is the annotation value of the scattering distribution, ${\eta_c^\gamma}$ is the albedo, $\gamma\geq1$ is the hyperparameter, and $c$ is the RGB color channel. Using logarithmic transformation, the radiation range can be greatly compressed, on the one hand, it can speed up the training, and on the other hand, it can reduce the problem of bright mutations caused by high-frequency phase functions. By introducing the hyperparameter $\gamma=4$, when $\gamma>0$, the input data will be close to normal. distributed. total loss function:
\begin{equation}\mathcal{L}_{total}=argmin\frac1N{\sum_{i=1}^N\mathcal{L}_S}\end{equation}
$N$ is the size of the annotation dataset. 
\begin{equation}\mathbb{Z}=\{(\mathcal{F}_1,\mathcal{F}(u_1,l_1)),(\mathcal{F}_2,\mathcal{F}(u_2,l_2)),...,(\mathcal{F}_N,\mathcal{F}(u_N,l_N))\}\end{equation}
$N$ is the size of the annotated data set, $\mathbb{Z}$ is the annotated data set.

\subsubsection{Attention module}
The attention module is implemented in many ways, such as ECANet (Efficient Channel Attention for Deep CNN), SKNet (Selective Kernel Networks), CBAM (Convolutional Block Attention Module). We use the Squeeze-and-Excitation Network (SENet) here.
Squeeze module. for the i-th layer input feature $\sum\mathrm{H}_i^\tau $, construct an 8-dimensional vector.
\begin{equation}v_i=\left\{B_i^p,B_i^f,B_i^T,f_i^\rho,f_i^f,f_i^T,g,\alpha\right\}\end{equation}

${v_i}$ represents the 8-dimensional input vector constructed from the input features of the i-th layer. $B_{i}^{\tau}=\frac1{n_{i}}\sum_{j=1}^{n_{i}}H_{i,j}^{\tau}$, $\mathfrak{f}_i^\tau=max{H}_{i,j}^\tau $. 
$B_{i}^{\tau}$ represents average pooling, $\mathfrak{f}_i^\tau$ represents maximum pooling. 
$n_i$ represents the total number of points in the i-th layer template,, and  represents the features of the j-th template point in the i-th layer. $\tau\in\{\rho,f,\mathcal{T}\}$,  where $\rho$ is the density, $f$ is the phase function, $\mathcal{T}$ is the transmission field, $\text{g}$ is the parameter of the Henyey-Greenstein phase function, and $\alpha=\cos^{-1}\left(\omega,l\right)$ is the angle between the viewing angle and the light direction. 

Weight coefficient. $w_i^\tau$ is the weight coefficient of the i-th layer template, $\mathcal{R}(v_i)$ is the vector obtained through the extrusion module, $\mathcal{R}(v_i)=\max{(v_i,0)}$ is the ReLU activation function, and $S(\mathcal{R}(v_i))=\frac1{1+e^{-\mathcal{R}(v_i)}}$ is the Sigmoid function.
\begin{equation}w_i^\tau=\mathcal{S}(\mathcal{R}(v_i))\end{equation}

Incentive module. $\overline{\sum H_i^\tau}$ is the self-attention feature after SENet.
\begin{equation}\overline{\sum H_i^\tau}=w_i^\tau\cdot\sum H_i^\tau \end{equation}





\begin{thebibliography}{00}
\bibitem{b1} W. Jarosz, M. Zwicker, H. W. Jensen, The beam radiance estimate for volumetric photon mapping, in: ACM SIGGRAPH 2008 classes, 2008, pp. 1–112.
\bibitem{b2} X. Deng, S. Jiao, B. Bitterli, W. Jarosz, Photon surfaces for robust, unbiased volumetric density estimation, ACM Transactions on Graphics 38 (4).
\bibitem{b3} S. Herholz, Y. Zhao, O. Elek, D. Nowrouzezahrai, H. P. Lensch, J. Kˇriv´anek, Volume path guiding based on zero-variance random walk theory, ACM Transactions on Graphics (TOG) 38 (3) (2019) 1–19. 1
\bibitem{b4} S. Kallweit, T. M¨uller, B. Mcwilliams, M. Gross, J. Nov´ak, Deep scattering: Rendering at mospheric clouds with radiance-predicting neural networks, ACM Transactions on Graphics (TOG) 36 (6) (2017) 1–11.1
\bibitem{b5} J. T. Kajiya, B. P. Von Herzen, Ray tracing volume densities, ACM SIGGRAPH computer graphics 18 (3) (1984) 165–174. 
\bibitem{b6} E. P. Lafortune, Y. D. Willems, Rendering participating media with bidirectional path tracing, in: Rendering Techniques’ 96: Proceedings of the Eurographics Workshop in Porto, Portugal, June 17–19, 1996 7, Springer, 1996, pp. 91–100. 8
\bibitem{b7} M. Pauly, T. Kollig, A. Keller, Metropolis light transport for participating media, in: Rendering Techniques 2000: Proceedings of the Eurographics Workshop in Brno, Czech Republic, June 26–28, 2000 11, Springer, 2000, pp. 11–22.8
\bibitem{b8} I. Georgiev, J. Krivanek, T. Hachisuka, D. Nowrouzezahrai, W. Jarosz, Joint importance sampling of low-order volumetric scattering., ACM Trans. Graph. 32 (6) (2013) 164–1.8
\bibitem{b9} C. Kulla, M. Fajardo, Importance sampling techniques for path tracing in participating media, in: Computer graphics forum, Vol. 31, Wiley Online Library, 2012, pp. 1519–1528.8
\bibitem{b10} J. Nov´ak, D. Nowrouzezahrai, C. Dachsbacher, W. Jarosz, Virtual ray lights for rendering scenes with participating media, ACM Transactions on Graphics (TOG) 31 (4) (2012) 1–11. 8
\bibitem{b11} M. Raab, D. Seibert, A. Keller, Unbiased global illumination with participating media, in: Monte Carlo and Quasi-Monte Carlo Methods 2006, Springer, 2006, pp. 591–605.8
\bibitem{b12} W. Jarosz, D. Nowrouzezahrai, I. Sadeghi, H. W. Jensen, A comprehensive theory of volumetric radiance estimation using photon points and beams, ACM transactions on graphics (TOG) 30 (1) (2011) 1–19.8
\bibitem{b13} H. W. Jensen, P. H. Christensen, Efficient simulation of light transport in scenes with participating media using photon maps, in: Seminal Graphics Papers: Pushing the Boundaries, Volume 2, 2023, pp. 301–310.8
\bibitem{b14} J. Kˇriv´anek, I. Georgiev, T. Hachisuka, P. V´evoda, M. Sik, D. Nowrouzezahrai, W. Jarosz, Unifying points, beams, and paths in volumetric light transport simulation, ACM Transactions on Graphics (TOG) 33 (4) (2014) 1–13.8
\bibitem{b15} B. Bitterli, W. Jarosz, Beyond points and beams: Higher-dimensional photon samples for volumetric light transport, ACM Transactions on Graphics (TOG) 36 (4) (2017) 1–12. 8
\bibitem{b16} P. Kutz, R. Habel, Y. K. Li, J. Nov´ak, Spectral and decomposition tracking for rendering heterogeneous volumes, ACM Transactions on Graphics (TOG) 36 (4) (2017) 1–16. 8
\bibitem{b17} Y. Yue, K. Iwasaki, B.-Y. Chen, Y. Dobashi, T. Nishita, Toward optimal space partitioning for unbiased, adaptive free path sampling of inhomogeneous participating media, in: Computer Graphics Forum, Vol. 30, Wiley Online Library, 2011, pp. 1911–1919. 9
\bibitem{b18} Y. Yue, K. Iwasaki, B.-Y. Chen, Y. Dobashi, T. Nishita, Unbiased, adaptive stochastic sampling for rendering inhomogeneous participating media, ACM Transactions on Graphics (TOG) 29 (6) (2010) 1–8.00
\bibitem{b19} J. Stam, Multiple scattering as a diffusion process, in: Rendering Techniques’ 95: Proceedings of the Eurographics Workshop in Dublin, Ireland, June 12–14, 1995 6, Springer, 1995, pp. 41–50.0
\bibitem{b20} D. Koerner, J. Portsmouth, F. Sadlo, T. Ertl, B. Eberhardt, Flux-limited diffusion for multiple scattering in participating media, in: Computer Graphics Forum, Vol. 33, Wiley Online Library, 2014, pp. 178–189.0
\bibitem{b21} C. Donner, H. W. Jensen, Light diffusion in multi-layered translucent materials, ACM Transactions on Graphics (ToG) 24 (3) (2005) 1032–1039.0
\bibitem{b22} J. R. Frisvad, T. Hachisuka, T. K. Kjeldsen, Directional dipole model for subsurface scattering, ACM Transactions on Graphics (TOG) 34 (1) (2014) 1–12. 8
\bibitem{b23} H. Wann Jensen, S. R. Marschner, M. Levoy, P. Hanrahan, A practical model for subsurface light transport, in: Seminal Graphics Papers: Pushing the Boundaries, Volume 2, 2023, pp. 319–326.8
\bibitem{b24} Y. Bengio, A. Courville, P. Vincent, Representation learning: A review and new perspectives, IEEE transactions on pattern analysis and machine intelligence 35 (8) (2013) 1798–1828.8
\bibitem{b25} Y. LeCun, Y. Bengio, G. Hinton, Deep learning, nature 521 (7553) (2015) 436–444.
\bibitem{b26} K. He, X. Zhang, S. Ren, J. Sun, Deep residual learning for image recognition, in: Proceedings of the IEEE conference on computer vision and pattern recognition, 2016, pp. 770–778.
\bibitem{b27} K. Simonyan, A. Zisserman, Very deep convolutional networks for large-scale image recognition, arXiv preprint arXiv:1409.1556.
\bibitem{b28} Y. Wu, M. Schuster, Z. Chen, Q. V. Le, M. Norouzi, W. Macherey, M. Krikun, Y. Cao, Q. Gao, K. Macherey, et al., Google’s neural machine translation system: Bridging the gap between human and machine translation, arXiv preprint arXiv:1609.08144.
\bibitem{b29} A. v. d. Oord, S. Dieleman, H. Zen, K. Simonyan, O. Vinyals, A. Graves, N. Kalchbrenner, A. Senior, K. Kavukcuoglu, Wavenet: A generative model for raw audio, arXiv preprint arXiv:1609.03499.
\bibitem{b30} A. Van Den Oord, N. Kalchbrenner, K. Kavukcuoglu, Pixel recurrent neural networks, in: International conference on machine learning, PMLR, 2016, pp. 1747–1756.
\bibitem{b31} O. Nalbach, E. Arabadzhiyska, D. Mehta, H.-P. Seidel, T. Ritschel, Deep shading: convolutional neural networks for screen space shading, in: Computer graphics forum, Vol. 36, Wiley Online Library, 2017, pp. 65–78.
\bibitem{b32} S. Bako, T. Vogels, B. McWilliams, M. Meyer, J. Nov´ak, A. Harvill, P. Sen, T. Derose, F. Rousselle, Kernel-predicting convolutional networks for denoising monte carlo renderings., ACM Trans. Graph. 36 (4) (2017) 97–1.
\bibitem{b33} C. Reddy, A. Chaitanya, A. Kaplanyan, C. Schied, M. Salvi, Interactive reconstruction of noisy monte carlo image sequences using a recurrent autoencoder, ACM Trans. Graph. 36 (4).
\bibitem{b34} M. Chu, N. Thuerey, Data-driven synthesis of smoke flows with cnn-based feature descriptors, ACM Transactions on Graphics (TOG) 36 (4) (2017) 1–14.
\bibitem{b35} D. Vicini, V. Koltun, W. Jakob, A learned shape-adaptive subsurface scattering model, ACM Transactions on Graphics (TOG) 38 (4) (2019) 1–15.
\bibitem{b36} L. Leonard, K. Hoehlein, R. Westermann, Learning multiple-scattering solutions for spheretracing of volumetric subsurface effects, in: Computer Graphics Forum, Vol. 40, Wiley Online Library, 2021, pp. 165–178.
\bibitem{b37} B. Mildenhall, P. P. Srinivasan, M. Tancik, J. T. Barron, R. Ramamoorthi, R. Ng, Nerf: Representing scenes as neural radiance fields for view synthesis, Communications of the ACM 65 (1) (2021) 99–106.
\bibitem{b38} T. M¨uller, F. Rousselle, J. Nov´ak, A. Keller, Real-time neural radiance caching for path tracing, arXiv preprint arXiv:2106.12372.
\bibitem{b39} M. Panin, S. Nikolenko, Faster rpnn: Rendering clouds with latent space light probes, in: SIGGRAPH Asia 2019 Technical Briefs, 2019, pp. 21–24. 



\end{thebibliography}
\end{document}